\documentstyle[prd,aps,eqsecnum,preprint,tighten]{revtex}
\def\ha{{1\over 2}}
\def\lan{\langle}
\def\ran{\rangle}
\def\bra#1{\lan#1|}
\def\ket#1{|#1\ran}
\def\del{\partial}
\def\normord#1{\mathopen{\hbox{\bf:}}#1\mathclose{\hbox{\bf:}}}
\def\frac#1,#2{{#1\over #2}}

\def\fr#1,#2{{#1\over #2}}

\begin{document}
\draft

\preprint{\vbox{\hfill SMUHEP/98--04}}

\title{The Mass Operator in the Light-Cone Representation}

\author{Gary McCartor}
\address{Department of Physics, Southern Methodist University, Dallas,
TX 75275}

\maketitle

\begin{abstract}
I argue that for the case of fermions with nonzero bare mass there is
a term in the matter density operator in the light-cone representation
which has been omitted from previous calculations.  The new term
provides agreement with previous results in the equal-time
representation for mass perturbation theory in the massive Schwinger
model.  For the DLCQ case the physics of the new term can be
represented by an effective operator which acts in the DLCQ subspace,
but the form of the term might be hard to guess and I do not know how
to determine its coefficient from symmetry considerations.

\end{abstract}


\section{Introduction}

In the light-cone representation, if the Fermi field has nonzero bare
mass the field $\psi_-$ is usually treated as a pure constraint.  That
is, the equation of motion for $\psi_-$ --- which contains no
$x^+$-derivatives --- is solved to write $\psi_-$ in terms of the true
degrees of freedom.  Solving the equation of motion requires the use
of boundary conditions however, and I shall argue in this letter that
a careful analysis shows that the boundary conditions which should be
chosen require additional degrees of freedom and that these degrees of
freedom lead to a term in the mass operator which has been left out of
--- as far as I am aware --- all previous light-cone calculations.

That the $\psi_-$ field contains degrees of freedom for the case of
zero bare mass has been known for some time\cite{mcc88,mcc91,mcc94};
the case of free fields is enough to show that.  Here I shall argue
that the case of nonzero bare mass is not qualitatively different
(regarding this point, see also Refs. \cite{mcc96,rob96}).  I shall
consider the case of the massive Schwinger model.  The existence of an
exact solution for the massless case and a large body of results in
the equal-time representation for the massive case makes the argument
particularly sharp.  In the case of the continuum, and with the new
term present, simple mass perturbation theory gives results in
agreement with mass perturbation theory in the equal-time
representation \cite{cjs75,col76,bsk76} and for surfaces ``near'' the
light-cone \cite{fpv96}.  For the case of periodicity conditions on
the characteristic (DLCQ)\cite{may76,pab85} there is a disagreement
with the older calculations.  I shall argue that the noncovariant
regulation of the singularity at $p^+ = 0$ leads to a composite
operator renormalization.  I believe that this last case provides a
useful example for those who wish to devise effective theories which
act only in the DLCQ subspace\cite{wir95,bur96}.

\section{The Schwinger Model}

The Schwinger model is electrodynamics of massless fermions in 1+1
dimensions \cite{sch62}.  In light-cone gauge the operator solution
is\cite{bmnnv}:
\begin{equation}
     \psi_+ = Z_+ e^{\Lambda_+^{(-)}}\sigma_+ e^{\Lambda_+^{(+)}}
\end{equation}
\begin{equation}
      \Lambda_+ = -i2\sqrt{\pi}(\tilde{\eta}(x^+) +
\tilde{\Sigma}(x^+,x^-))
\end{equation}
\begin{equation}
       Z_+^2 = \fr{m^2e^\gamma},{8\pi\chi}
\end{equation}
\begin{equation}
     \psi_- = Z_-e^{\Lambda_-^{(-)}}\sigma_- e^{\Lambda_-^{(+)}}
\label{pmc}
\end{equation}
\begin{equation}
        Z_-^2 = \fr{\chi e^\gamma},{2\pi}
\end{equation}
\begin{equation}
      \Lambda_- = -i2\sqrt{\pi}\phi(x^+)
\end{equation}
\begin{equation}
        A^- = \fr{1},{m} \partial_+ (\tilde{\eta} + \tilde{\Sigma})
\end{equation}
\begin{equation}
        A^+ = 0
\end{equation}
Here, $\phi(x^+)$ is the left-moving component of a free, massless
scalar field while $\tilde{\eta}(x^+)$ is the left-moving component of
a free, massless ghost field, both regulated in the infrared with a
Klaiber\cite{kla67} regulator of parameter $\chi$.  $\tilde{\Sigma}$
is a free massive psuedoscalar field of mass $e/\sqrt{\pi}$.
$\sigma_+$ and $\sigma_-$ are spurions.  The vacuum must be chosen to
be a theta-state, formed as:
\begin{equation}
             \ket{\Omega(\theta)} \equiv \sum_{n=-\infty}^\infty
e^{iM\theta}\ket{\Omega(M)} \quad;\quad \ket{\Omega(M)} =
({\sigma}_+^*{\sigma}_-)^M \ket{0}
\end{equation}
The physical subspace is formed by applying all polynomials in
$\tilde{\Sigma}$ to $\ket{\Omega(\theta)}$.  A detailed discussion of
this solution is given elsewhere \cite{bmnnv} but the most important
point for the present discussion is (\ref{pmc}): the field $\psi_-$ is
not zero but is isomorphic to the left moving component of a free,
massless Fermi field --- it contains degrees of freedom.  We also note
that:
\begin{equation}
     \bra{\Omega(\theta)} \bar{\psi} \psi \ket{\Omega(\theta)} =
-\fr{m},{2\pi} e^\gamma cos\theta
\end{equation}

If we now consider the case of imposing periodicity conditions on the
characteristics, in particular: $\psi_+(x^- + 2L) = -\psi_+(x^-)$;
$\psi_-(x^+ + 2L) = -\psi_-(x^+)$, we find the solution \cite{mcc94}:
\begin{equation}
\psi_+ = {1\over \sqrt{2L}}
e^{-\lambda_+^*(x)}
\sigma_+(x)
e^{\lambda_+(x)}
\end{equation}
\begin{equation}
\lambda_+(x) = -i\sqrt{{\pi\over L}}\sum_{n=1}^\infty
{1\over \sqrt{p_-(n)}} C(n)e^{-ip(n)x}      
\end{equation}
\begin{equation}
       p_-(n) = {{n\pi}\over{L}}\quad ;\quad p_+(n) =
{{m^2L}\over{4n\pi}}
\end{equation}
\begin{equation}
\psi_- = {1\over\sqrt{2L}} e^{-\lambda_D^*(x^+)}
\sigma_-(x)e^{\lambda_D(x^+)}
\label{pmp}        
\end{equation}
\begin{equation}
\lambda_D(x^+) = \sum_{n=1}^\infty{1\over\sqrt{n}} D(n)e^{-ik_+(n)x^+}
\end{equation}
\begin{equation}
\sigma_+(x) = e^{-i{\sqrt{\pi}\over 4Lm}\left(Q_+(x^--x^+)\right)}
\sigma_+(0)e^{-i{\sqrt{\pi}\over 4Lm}\left(Q_+(x^--x^+)\right)}
\end{equation}
\begin{equation}
\sigma_-(x) = e^{-i{\sqrt\pi\over 4Lm}\left(Q_-(x^+-x^-)\right)}
\sigma_-(0)e^{-i{\sqrt\pi\over 4Lm}\left(Q_-(x^+-x^-)\right)}
\end{equation}
\begin{equation}
A^-=-{i\over \sqrt{L}m}\sum_{n=1}^\infty{p^-(n)
\over \sqrt{p_-(n)}}\left( C(n)e^{-ip(n)x}-
 C^*(n)e^{ip(n)x}\right) 
-{1\over Lm^2}Q_+
\end{equation}
\begin{equation}
   A^+=-{1\over Lm^2}Q_-       
\end{equation}
Here, the $C$'s are the fusion operators associated with bosonizing
the $\psi_+$ field and the $D$'s are the fusion operators associated
with bosonizing the $\psi_-$ field. The vacuum must be chosen to be a
theta-state, formed as:
\begin{equation}
             \ket{\Omega(\theta)} \equiv \sum_{n=-\infty}^\infty
e^{iM\theta}\ket{\Omega(M)} \quad;\quad \ket{\Omega(M)} =
({\sigma}_+^*{\sigma}_-)^M \ket{0}
\end{equation}
The physical subspace is formed by applying all polynomials in $C^*$'s
to $\ket{\Omega(\theta)}$.  A detailed discussion of this solution is
given elsewhere \cite{mcc94}, but the most important point for the
present discussion is (\ref{pmp}): the field $\psi_-$ is not zero; it
is not quite isomorphic to a free, massless Fermi field due to the
fact that the $\sigma_-$ operator picks up a dependence on $x^-$, but
the degrees of freedom are the same as for the left moving component
of a free, massless Fermi field.  We also note that:
\begin{equation}
     \bra{\Omega(\theta)} \bar{\psi} \psi \ket{\Omega(\theta)} = -
\fr{1},{L} cos\theta
\end{equation}
The chiral condensate goes to zero as the periodicity length, $L$,
goes to infinity, which is not the result in the continuum solution
presented above \cite{mcc97}.

\section{The Massive Schwinger Model}

Now let us consider adding a mass term, that is, we take the
Lagrangian to be:
\begin{equation}
{\cal L} = \ha \left(i \bar{\psi} \gamma^{\mu} \partial_{\mu} \psi - i
\partial_{\mu} \bar{\psi} \gamma^{\mu} \psi \right) - \fr{1},{4}
F^{\mu \nu} F_{\mu \nu} - A^{\mu} {J^\prime}_{\mu} - \mu\bar{\psi}\psi
\end{equation}
The equations of motion are:
\begin{equation}
\fr{\partial \psi_+},{\partial x^+} + i{1\over4}e(A^-\psi_++\psi_+A^-)
+ i\ha\mu\psi_- = 0
\end{equation}
\begin{equation}
{\del\psi_-\over\del x^-}+i{1\over4}e(A^+\psi_-+\psi_-A^+) +
i\ha\mu\psi_+ =0
\label{pmeom} 
\end{equation}
\begin{equation}
\fr{\partial^2 A^-},{\partial x^{-2}} = -\ha {J}^+
\end{equation}
\begin{equation}
-\fr{\partial^2 A^+},{\partial x^{+2}} + \fr{\partial^2 A^-},{\partial
x^+ \partial x^-} = \ha {J}^-
\end{equation}
The equation for $\psi_-$, (\ref{pmeom}), contains no $x^+$
derivatives.  If we take the continuum case, where $A^+ = 0$ then we
may take:
\begin{equation}
   \psi_- = -\int i\ha\mu\psi_+ dx^-
\end{equation}
What has usually been done in the past (always, as far as I know) is
to define the antiderivative to be such that if:
\begin{equation}
    \psi_+(x^-) = {1\over\sqrt{2L}}\sum_{n=1}^\infty b(n)
   e^{-ik_-(n)x^-} + d^*(n) e^{ik_-(n)x^-}
\end{equation}
then:
\begin{equation}
    \int \psi_+(x^-) = {1\over\sqrt{2L}}\sum_{n=1}^\infty
   \fr{1},{-ik_-(n)}b(n) e^{-ik_-(n)x^-} + \fr{1},{ik_-(n)}d^*(n)
   e^{ik_-(n)x^-}
\end{equation}
{}From now on we shall understand the antiderivative to mean that
particular one.  But we can also take:
\begin{equation}
   \psi_- = \psi_-^0-\int i\ha\mu\psi_+ dx^-
\end{equation}
Where $\psi_-^0$ is any solution to the homogeneous equation, which in
this case means any function of $x^+$.  Looking at (\ref{pmc}) we see
that if the solution for nonzero $\mu$ is to go smoothly into the
Schwinger model we should not take $\psi_-^0$ to be zero but rather:
\begin{equation}
      \psi_-^0 = Z_-e^{\Lambda_-^{(-)}}\sigma_- e^{\Lambda_-^{(+)}}
\end{equation}
With that choice we find that in the physical subspace the operator
$P^-$ gets a correction
%
%
%
\begin{eqnarray}
\delta P^- &=& \mu Z_-Z_+\int_{-\infty}^\infty
\left(\sigma_-^*\sigma_+
\normord{e^{-i\sqrt{\pi}\tilde{\Sigma}(0,x^-)}} + C.C.\right) dx^-
\nonumber\\ 
& &\qquad+ \ha\mu^2 \int_{-\infty}^\infty \left(\psi_+^*(0,x^-)
\left[-\int i\ha \psi_+ \right] + C.C.\right)dx^-
\label{pmt}
\end{eqnarray}
The first term depends on the vacuum angle, $\theta$, while the second
term does not.  Furthermore, the second term needs regularization ---
a very important issue, but one that need not concern us here.  One
might expect a factor of one-half in front of the first term, but the
equation is correct as it stands as can be verifyed by checking the
Heisenberg equation for the $x^+$-derivative of $\psi_+$.  The
constants $Z_-$ and $Z_+$ are wavefunction renormalization constants;
they depend on $\mu$ and are determined by the requirement that the
fields remain canonically normalized.  We know from the Schwinger
model solution that:
\begin{equation}
        Z_- = Z_+(\mu) = \sqrt{\fr{\chi e^\gamma},{2\pi}} \;+\; {\cal
O}(\mu)
\end{equation}
\begin{equation}
           Z_+ = Z_+(\mu) = \sqrt{\fr{m^2e^\gamma},{8\pi\chi}} \;+\;
{\cal O}(\mu)
\end{equation}
If we perform a straightforward first order perturbation calculation
for the state of one Schwinger particle:
\begin{equation}
        \ket{p} \equiv \tilde{\Sigma}^*(p)\ket{\Omega(\theta)}
\end{equation}
only the first term contributes and we get:
\begin{equation}
\delta M^2 = \bra{p} P^+ \delta P^- \ket{p} = 2 m \mu e^\gamma
cos\theta
\end{equation}
This result is in agreement with previous calculations done in the
equal-time representation, both in the continuum \cite{cjs75,col76}
and with periodicity conditions\cite{bsk76,nak83}, and in a
representation using a periodic surface ``near'' the light-cone
\cite{fpv96}.  Indeed, for the case of the equal-time, continuum
calculations, it is easy to evaluate the fields on the characteristics
(to first order in $\mu$) and see that, not only are the results
numerically equal, but the operators are the same and so the
calculations are completely equivalent \cite{mrv97}.

If we now turn to the DLCQ case, by the same argument we would choose:
\begin{equation}
\psi_-^0 = {1\over\sqrt{2L}} e^{-\lambda_D^*(x^+)}
\sigma_-(x)e^{\lambda_D(x^+)}
\end{equation}
That will lead to a new term in $P^-$ operator which, in the physical
subspace is given by:
\begin{equation}
       \delta P^- \subset \mu \sigma_-^*\sigma_+ \int_{-L}^L {1\over
{2L}} e^{\lambda_+^{(-)}(0,x^-)} e^{\lambda_+^{(+)}(0,x^-)} dx^- +
C.C.
\end{equation}
\begin{equation}
\lambda_+(x) = -i\sqrt{{\pi\over L}}\sum_{n=1}^\infty {1\over
\sqrt{p_-(n)}}\left( C(n)e^{-ip(n)x} + C^*(n)e^{ip(n)x}\right)
\end{equation}
Note that we may replace the operator $\sigma_-^*\sigma_+$ by
$e^{i\theta}$ and similarly for the complex conjugate, and thus have
an operator which operates entirely within the DLCQ subspace --- an
important point from the point of view of effective theories.

The state of one Schwinger particle is:
\begin{equation}
        \ket{n} \equiv C^*(n)\ket{\Omega(\theta)}
\end{equation}
And mass perturbation theory gives:
\begin{equation}
         \bra{n} P^+ \delta P^- \ket{n} = 4 \pi \mu \fr{1},{L}
cos\theta
\end{equation}
The mass shift goes to zero as $L$ goes to infinity for the same
reason that the chiral condensate goes to zero as $L$ goes to
infinity.

Introducing periodicity conditions on $x^+ = 0$ regulates the
singularity at $p^+ = 0$ in a very abrupt way which violates the
Poincar\'e symmetries, particularly parity.  Thus we should not expect
to absorb all the infinities into mass, wavefunction and coupling
renormalizations.  The operator mixes with itself under
renormalization and we should write:
\begin{equation}
       \delta P^- \subset Z\mu \sigma_-^*\sigma_+ \int_{-L}^L {1\over
 {2L}} e^{\lambda_+^{(-)}(0,x^-)} e^{\lambda_+^{(+)}(0,x^-)} dx^- +
 C.C.
\end{equation}
where $Z$ is {\it not} determined by requiring canonical normalization
of the fields but must be determined by fitting to data.  Taking the
continuum calculation as data we write:
\begin{equation}
        Z = Z(\mu) = \fr{L m e^\gamma},{2\pi} \;+\; {\cal O}(\mu)
\end{equation}
While this should perhaps be regarded as a conjecture at the moment,
it is testable within the context of the model: having chosen $Z$ to
fit the mass we can then unambiguously predict the S-matrix.  Since
the massive Schwinger model has a nontrivial S-matrix, we have a
nontrivial test of the conjecture.  I hope to report results on that
test in the near future.

\section{Effective Theories}

I believe the example we have been considering contains at least three
important lessons for those who wish to make effective theories which
act in the DLCQ subspace and incorporate the effects of zero modes
with additional operators \cite{wir95,bur96}.  The first is good news:
the effects of zero modes we have been considering here can in fact be
incorporated into effective operators. But the second is that the form
of the operators may be hard to guess, particularly if one thinks
entirely within the DLCQ framework.  Since the $\psi_+$ field is
irreducible in the DLCQ subspace we are guaranteed to be able to write
our effective operator in terms of it; indeed we can and it seems
rather natural if we bosonize the field and think of zero modes.  As a
functional of $\psi_+$ it is:
\begin{eqnarray}
Z\mu e^{-i\theta} {1\over {2L}} \int_{-L}^L dx^-
{\bf :}\lim_{y^-\rightarrow
0}&e&^{\left(2i\pi\int \left[\normord{\psi_+^*(y^-)\psi_+(y^-)} -
{1\over2L}\int_{-L}^L\normord{\psi_+^*(z^-)
\psi_+(z^-)}dz^-\right]\right)^{(-)}}\psi_+(y^-)\\
\times &e&^{\left(2i\pi\int
\left[\normord{\psi_+^*(y^-)\psi_+(y^-)} - {1\over
2L}\int_{-L}^L\normord{\psi_+^*(z^-)\psi_+(z^-)}dz^-\right]\right)^{(+)}}
\psi_+^*(x^-){\bf :} + C.C.\nonumber
\end{eqnarray}
While this operator can certainly be found I think it may not be the
first thing that would pop into everyone's head.

The final point is that even if the form of the operator is given, the
coefficient cannot always be found by restoring the broken symmetry.
Here the relevant symmetry is parity but the effective operator does
not restore parity.  One can, in fact, find a similar operator which
does restore parity (and gives the correct limit of the chiral
condensate) but it does not act in the DLCQ subspace\cite{mcc91}.  It
requires a great enlargement of the DLCQ space in a way that breaks
the periodicity conditions.  I believe that the operator we have been
studying here has an analog in four dimensions and that the analog
operator probably plays a role in chiral symmetry breaking.  I do not
yet know what the form of the operator is and I expect that the
coefficient will have to be fit to data.

\acknowledgments
\noindent
I appreciate the efforts of Dave Robertson and Eliana Vianello who
helped me understand the factors in (\ref{pmt}).  This work was
supported in part by grants from the U.S. Department of Energy.

\end{document}